# Simulation and experimental studies of induction hardening behavior of a new medium-carbon, low-alloy wear resistance steel

**Vahid Javaheri · John-Inge Asperheim · Bjørnar Grande · Tun Tun Nyo· David Porter**

**Abstract** Flux2D commercial software together with a Gleeble thermomechanical simulator has been employed to numerically and physically simulate the material properties profile of an induction hardened slurry transportation pipe made of a recently developed 0.4 wt.% C, Nb-microalloyed steel. After calculating the thermal history of a 400 mm diameter, 10 mm thick pipe at various positions through the thickness, different heating and cooling paths were physically simulated using the Gleeble machine to predict the through-thickness material microstructure and hardness profiles. The results showed that by coupling a phase transformation model considering the effect of heating rate on the austenite transformation temperatures which allows calculations for arbitrary cooling paths with calculated induction heating and quenching thermal cycles, it has been possible to design induction hardening parameters for a slurry transport pipe material.

Vahid Javaheri
Material engineering and production technology, University of Oulu, Oulu, Finland
E-mail: vahid.javaheri@oulu.fi

John_Inge Asperheim
R&D, EFD Induction a.s., Skien, Norway
E-mail: Joh-Inge.Asperheim@efd-induction.com

Bjørnar Grande
R&D, EFD Induction a.s., Skien, Norway
E-mail: Bjornar.Grande@efd-induction.com

Tun Tun Nyo
Material engineering and production technology, University of Oulu, Oulu, Finland
E-mail: Tun.Nyo@oulu.fi

David Porter
Material engineering and production technology, University of Oulu, Oulu, Finland
E-mail: David.porter@oulu.fi

## 1 Introduction

In the long-distance transportation of cement and mining slurry, pipeline transportation dominates because of its high efficiency and low cost as well as its environmental benefits. One way of meeting the demanding requirements for good performance of slurry pipelines is to have a gradient of hardness and toughness through the pipe wall thickness in order to combine good wear resistance and sufficient body toughness. To be able to achieve such properties, the following production route would be not only metallurgically but also economically advantageous: continuous casting, hot strip rolling and coiling, slitting into skelp, cold forming and high-frequency induction welding into pipe followed by induction heat treatment to give a hard-inside pipe surface combined with a tough pipe body. Amongst all the production steps, induction hardening has the predominant influence on the final properties and there is an essential need to have a good picture of the phase transformations occurring during the induction hardening, which requires precise evaluation of the temperature distribution during heating and cooling. Many studies have been carried out using numerical methods to understand and optimize the induction hardening of the pipes and tubes [1]-[4], but there are only a few works dealing with the effect of induction hardening parameters on microstructure and material properties. Material properties obtained by induction hardening vary through the thickness from the hardened surface toward the bulk as large changes in the peak temperature during the induction hardening cause a wide variety of phase transformation conditions.

Recently, Montalvo-Urquizo et al. [5] numerically investigated the induction hardening and phase transformations occurring through the thickness of a gearing component made of 42CrMo4. However, they reported that due to the high cooling rate in their designed system, most of the austenite was transformed to martensite and there was no gradient of material properties through the hardened depth.

Here, the main aim of this work is the coupling of phase transformation modelling based on data obtained from continuous cooling transformation (CCT) diagrams with predicted induction hardening thermal cycles in order to predict the gradient of material microstructure and hardness through the pipe body after quenching.

## 2 Problem Description

The problem is to predict the distribution of material properties in a specific steel, in the form of a 400 mm diameter pipe with a wall thickness of 10 mm, after scanning induction heating and subsequent rapid quenching, based on the predicted temperature history through the wall thickness. To have an accurate prediction, realistic thermal cycles for each layer of the wall thickness need to be combined with phase transformation modelling based on information from CCT diagram.

## 3 Simulation

In order to solve the problem and to simulate the whole process, a flexible combined electromagnetic - thermal calculation model for induction hardening has been applied using the commercially available finite element software Flux 2D. An axisymmetric 2D geometry has been considered for the model, as illustrated in Figure 1. The pipe is considered to move at a constant velocity of 10 mm/s past a stationary induction heating inductor and quenching system.

To analyze the magnetic fields and magnetic flux density distribution, Maxwell's equation with field strength and temperature dependent material properties has been solved in a steady state AC magnetic computation. The generated heat and temperature distribution through the pipe have been calculated by solving the governing heat transfer differential equations considering convection and radiation for both the heating and cooling parts. In the water-cooling section, the intensity of the convection is a function of the surface temperature of the pipe.

## 4 Experiments

A Gleeble 3800 thermomechanical simulator has been employed to physically simulate different thermal cycles, i.e. different heating and cooling rates and also to provide dilatometric phase transformation data required for developing a continuous cooling transformation (CCT) model based on the approach of Pohjonen [6]. The cylindrical Gleeble samples were cut in half perpendicular to their axis at the position of the control thermocouple used for monitoring the temperature during the thermal cycles. The starting material was laboratory hot rolled and directly water quenched to the quench-stop temperatures of 420 °C followed by furnace cooling to produce an essentially isothermal lower bainite starting microstructure. The composition of the studied steel together with the initial bulk mean hardness value and mean prior austenite grain size (PAGS) on through-thickness cross-sections parallel to the rolling direction are given in Table 1. More details about the material composition and as-rolled properties can be found in refs. [7],[8].

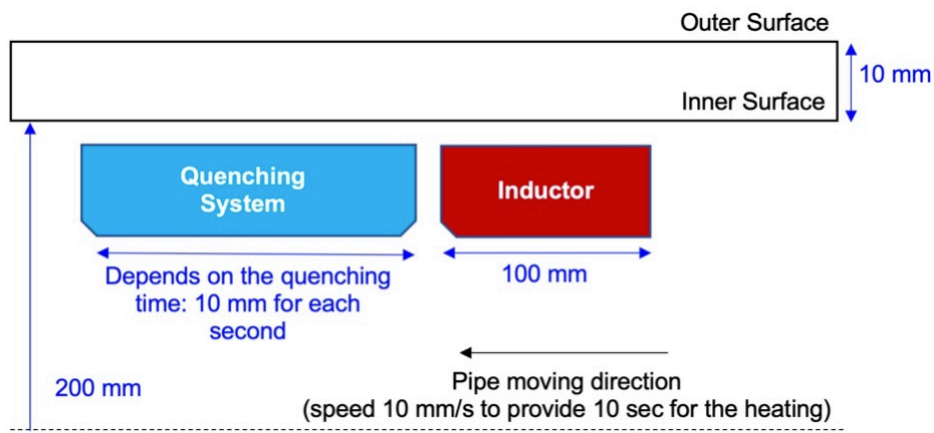

**Figure 1**. Geometry for the induction hardening simulation

**Table 1.** Composition of studied material along with its prior austenite grain size and the hardness

| Composition (wt.%) | | | | | | | | PAGS[1] ($\mu m$) | Hardness (HV 10) |
|---|---|---|---|---|---|---|---|---|---|
| C | Si | Mn | Cr | Ni | Mo | Nb | N | 29±5 | 415±5 |
| 0.40 | 0.19 | 0.24 | 0.92 | 0.02 | 0.48 | 0.013 | 0.004 | | |

1 Prior Austenite Grain Size expressed as mean equivalent circle diameter on a cross-section containing the rolling and normal directions.

## 5 Results and Discussion

### 5.1 On-heating Temperature Profile

The simulated temperature profile of every millimeter of pipe thickness as a function of time is presented in Figure 2(a). The aim was to reach a peak temperature of about 950° in 10 seconds at the inner surface with a gradual decrease in peak temperature toward the outer surface. As Figure 2(b) shows, in the very beginning of the heating process the heating rates at the inner layers are significantly higher than further inside the pipe until Curie temperature is reached. This is due to the high relative magnetic permeability and low penetration depth. At this stage the inner surface is heated directly by the induced current while regions further inside the pipe is heated by the thermal conduction. Above the Curie temperature the relative permeability drops to unity which increases the penetration depth and leads to a moderate heating rate [9].

Therefore, beside the highest heating rate, the highest peak temperature will also be reached at or just below the inner surface. In the other words, both induced joule heat and temperature are much higher on the inner surface than at the outer surface which leads to clear initial temperature differences and heating rates on the inner and outer wall layers. Later, due to the thermal conductivity of material, the heating rate converges to almost the same level of about 50 °C/s through the pipe wall thickness. Recently, the present authors [10] have shown that a heating rate of 50°C/s in the austenitisation region produces the finest prior austenite grain size. The peak temperature of every millimeter of pipe thickness obtained from the simulations of the designed heating process is given in Figure 3. It can be seen that almost 40% of wall thickness should be fully austenitized i.e. see temperatures above the austenitization finish temperature (red line), 25% is predicted to be heated into the two-phase region (between blue and red line) and 35% should remain below the lower critical temperature (the blue line), i.e. the tempered as-rolled microstructure. In this way, a smooth gradient of material properties is expected with hard inner layers and a softer, tougher outer body.

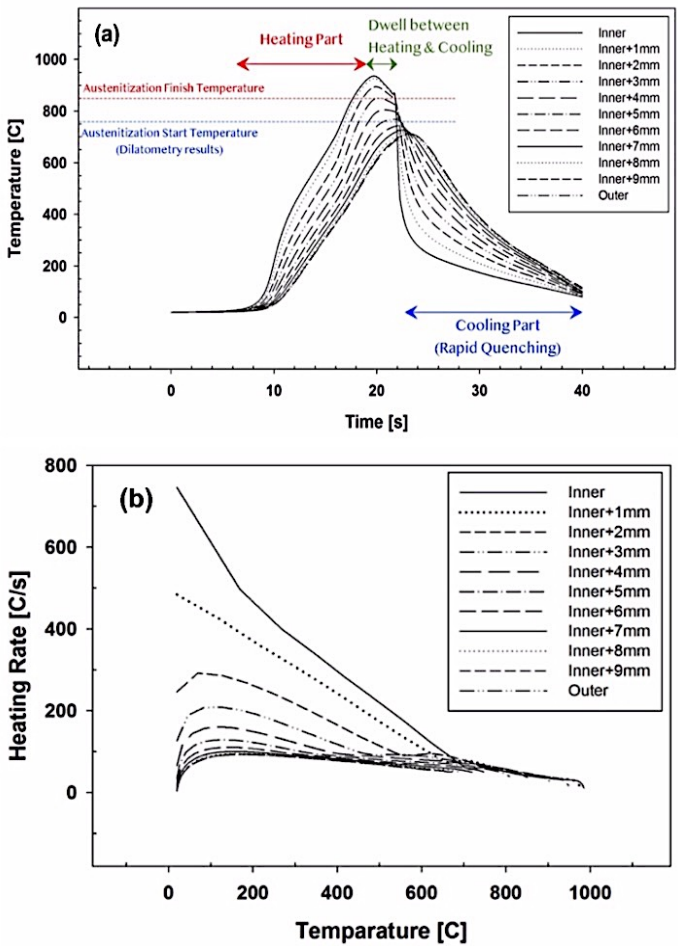

**Figure. 2** a) The temperature profile of the whole process including both heating and cooling, b) heating rate at the different thicknesses

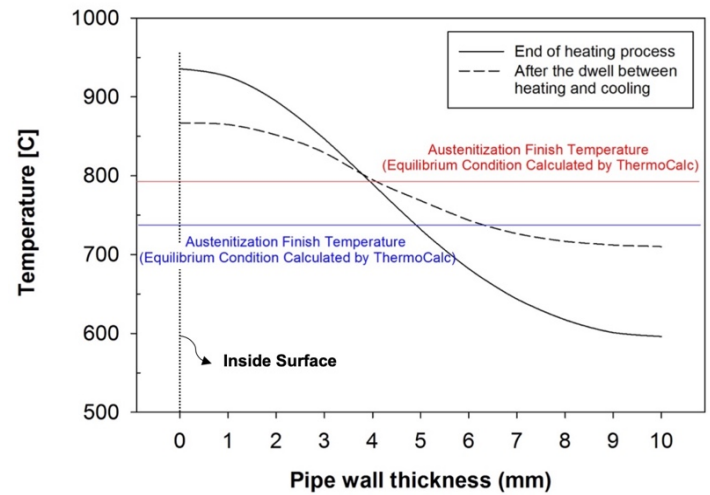

**Figure. 3** The peak temperature for the different thicknesses after heating and at the beginning of cooling

### 5.2 On-cooling Temperature Profile

Figure 4(a) shows the experimental results of the phase fractions of the studied material after heating to 950 °C and subsequent cooling at different rates. It can be seen that when austenitization is complete, almost 100% martensite can be achieved when the cooling rate is 60 °C/s or more while polygonal ferrite appears when the cooling rate is less than about 10 °C/s. To be able to estimate the phase fraction distribution for any continuous cooling path, not just linear cooling, an empirical model [6,11] has been built around the dilatometry results. The different cooling paths for every millimeter of pipe thickness resulting from water quenching is presented in Figure 4(b) together with the CCT diagram from the empirical model.

## 6 Results and Discussion

### 6.1 On-heating Temperature Profile

The simulated temperature profile of every millimeter of pipe thickness as a function of time is presented in Figure 2(a). The aim was to reach a peak temperature of about 950° in 10 seconds at the inner surface with a gradual decrease in peak temperature toward the outer surface. As Figure 2(b) shows, in the very beginning of the heating process the heating rates at the inner layers are significantly higher than further inside the pipe until Curie temperature is reached. This is due to the high relative magnetic permeability and low penetration depth. At this stage the inner surface is heated directly by the induced current while regions further inside the pipe is heated by the thermal conduction. Above the Curie temperature the relative permeability drops to unity which increases the penetration depth and leads to a moderate heating rate [9].

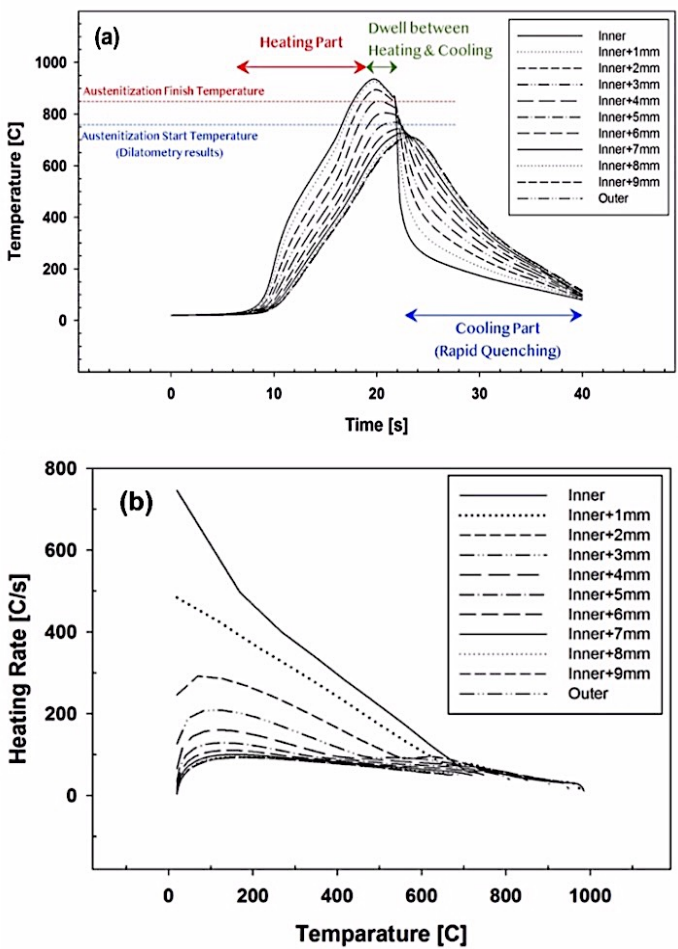

**Figure. 2** a) The temperature profile of the whole process including both heating and cooling, b) heating rate at the different thicknesses

Therefore, beside the highest heating rate, the highest peak temperature will also be reached at or just below the inner surface. In the other words, both induced joule heat and temperature are much higher on the inner surface than at the outer surface which leads to clear initial temperature differences and heating rates on the inner and outer wall layers. Later, due to the thermal conductivity of material, the heating rate converges to almost the same level of about 50 °C/s through the pipe wall thickness. Recently, the present authors [10] have shown that a heating rate of 50°C/s in the austenitisation region produces the finest prior austenite grain size. The peak temperature of every millimeter of pipe thickness obtained from the simulations of the designed heating process is given in Figure 3. It can be seen that almost 40% of wall thickness should be fully austenitized i.e. see temperatures above the austenitization finish temperature (red line), 25% is predicted to be heated into the two-phase region (between blue and red line) and 35% should remain below the lower critical temperature (the blue line), i.e. the tempered as-rolled microstructure. In this way, a smooth gradient of material properties is expected with hard inner layers and a softer, tougher outer body.

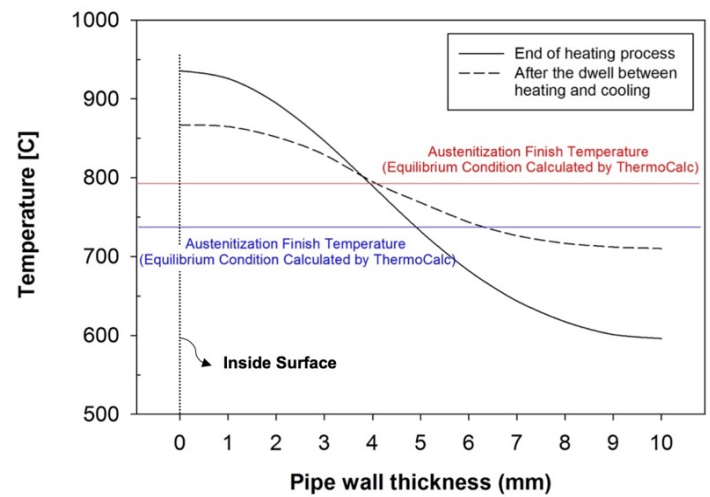

**Figure. 3** The peak temperature for the different thicknesses after heating and at the beginning of cooling

### 6.2 On-cooling Temperature Profile

Figure 4(a) shows the experimental results of the phase fractions of the studied material after heating to 950 °C and subsequent cooling at different rates. It can be seen that when austenitization is complete, almost 100% martensite can be achieved when the cooling rate is 60 °C/s or more while polygonal ferrite appears when the cooling rate is less than about 10 °C/s. To be able to estimate the phase fraction distribution for any continuous cooling path, not just linear cooling, an empirical model [6,11] has been built around the dilatometry results. The different cooling paths for every millimeter of pipe thickness resulting from water quenching is presented in Figure 4(b) together with the CCT diagram from the empirical model.

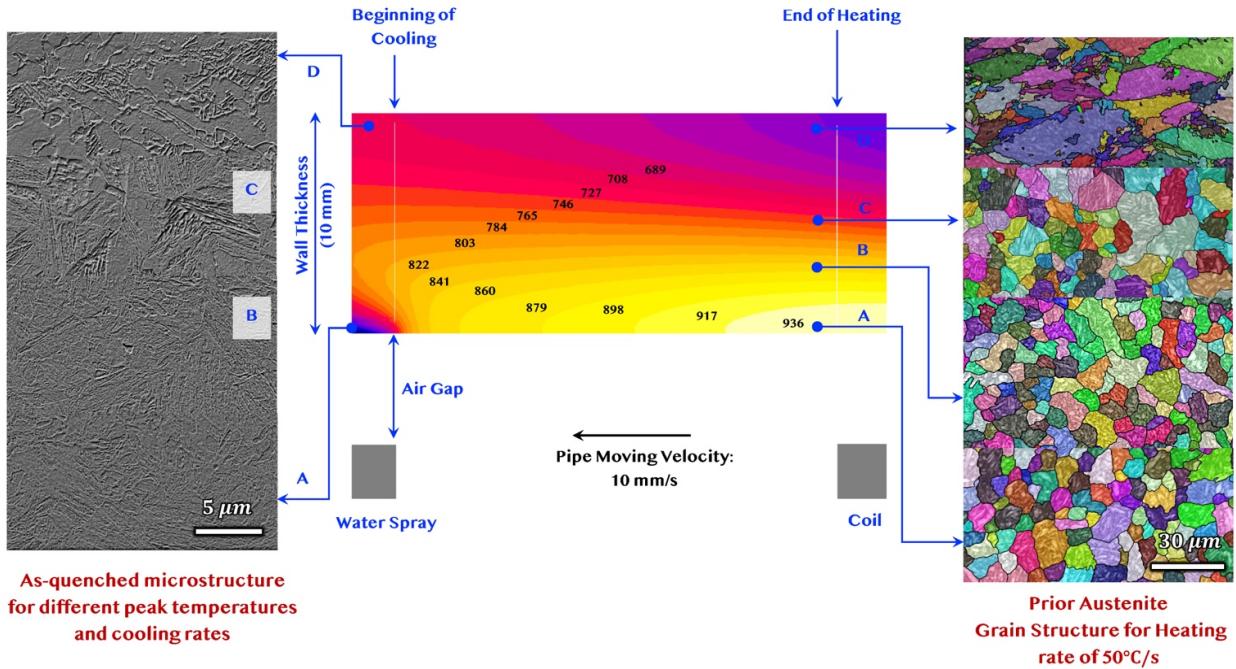

**Figure 5.** The temperature history of the pipe wall thickness during its travel between heating and cooling steps together with the reconstructed prior austenite grain structure and final microstructure after water spray for the highlighted positions of A, B, C, and D.

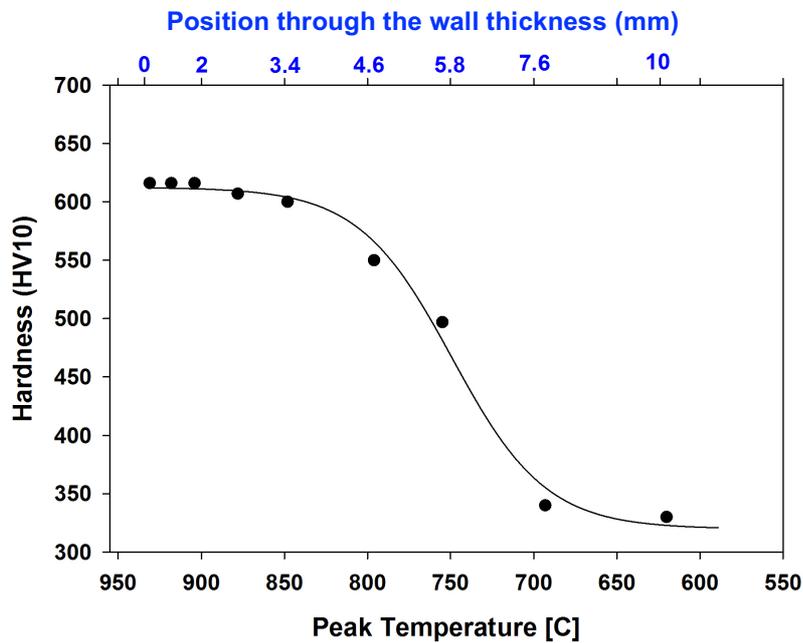

**Figure 6.** Hardness as a function of peak temperature for the different positions through the wall thickness after heating and cooling

## 7 Acknowledgments

The authors are grateful for financial support from the European Commission under grant number 675715 − MIMESIS − H2020-MSCA-ITN-2015, which is a part of the Marie Sklodowska-Curie Innovative Training Networks European Industrial Doctorate programme.